\documentclass[aps,prb,showpacs,superscriptaddress,twocolumn]{revtex4-1}
\usepackage{amsfonts}

\usepackage{amssymb}
\usepackage{amsmath}
\usepackage{graphicx}
\usepackage{hyperref}
\usepackage{siunitx}
\usepackage[dvipsnames]{xcolor}
\usepackage{nicematrix}
\usepackage[normalem]{ulem}
\usepackage[version=4]{mhchem}
\usepackage{float}
\usepackage{siunitx}
\usepackage[resetlabels,labeled]{multibib}
\newcites{App}{App Readings}

\hypersetup{colorlinks=true,
            linkcolor=blue,
            citecolor=blue,
            urlcolor=orange}

\newcommand{\I}{{\rm i}}

\begin{document}

\title{Frequency dependence of temporal spin stiffness and short-range magnetic order \\in the doped 
two-dimensional Hubbard model}
\author{I. A. Goremykin}
\affiliation{Center for Photonics and 2D Materials, Moscow Institute of Physics and Technology, Institutsky lane 9, Dolgoprudny
141700, Moscow region, Russia}
\author{A. A. Katanin}
\affiliation{Center for Photonics and 2D Materials, Moscow Institute of Physics and Technology, Institutsky lane 9, Dolgoprudny
141700, Moscow region, Russia}
\affiliation{M. N. Mikheev Institute of Metal Physics, Kovalevskaya Street 18,
Ekaterinburg {620219}, Russia.}

\begin{abstract}
We study doping and temperature dependencies of temporal and spatial spin stiffnesses of the Hubbard model within the mean field approach for incommensurate magnetic order. We show that the frequency dependence of temporal spin stiffness {within the considered mean field approach} is crucial to obtain small values of correlation length, comparable to those observed in cuprates. 
Using the obtained spin stiffnesses, we obtain the temperature and doping dependence of correlation length within the large-$N$ limit of the respective nonlinear sigma model. In agreement with the experimental data on La$_{2-x}$Sr$_x$CuO$_4$ we obtain short range magnetic order with relatively small correlation length at $0.1 \lesssim x\lesssim 0.2$, and magnetically ordered ground state in the narrow doping region $0.05\lesssim x \lesssim 0.1$. 
The latter state may correspond to the spin-frosen state, observed in the experimental data on La$_{2-x}$Sr$_x$CuO$_4$.
\end{abstract}


\maketitle


Magnetic properties of high-$T_c$ compounds remain an active research field, despite long time of their study. For the most studied high-$T_c$ compound La$_{2-x}$Sr$_x$CuO$_4$ the commensurate long-range magnetic order is destroyed already at small doping
(see, e.g., Refs. \onlinecite{Plakida,StaticOrd}), while the short-range magnetic order with the 
wave vector 
$\mathbf{Q}%
=(\pi,\pi-\delta)$  is formed with doping \cite{ExpQ1,ExpQ2,ExpQ3,StaticOrd,Plakida}. Interestingly, despite the presence of strong electronic correlations, for not too small dopings $x=1-n>0.03$ the correlation length is quite moderate \cite{ExpInel1,ExpInel2}, $\xi\sim (2$-$5)a$, except for the doping $x\simeq 0.12$ (where stripe appearance was suggested). {At the same time, the signs of static spin freezing and long-range magnetic order were observed at $x\simeq 0.12$} (see Refs. \onlinecite{LRO_in_cuprate_Katano2000,LRO_in_cuprate_Khaykovich2002,LRO_in_cuprate_Khaykovich2003,LRO_in_cuprate_Chang2008,LRO_in_cuprate_Khaykovich2005,LRO_in_cuprate_Lake2002_Nature,NMR1,NMR2,StrMet}). These observations show a strong change of magnetic properties with doping even in the moderately doped regime $x\gtrsim 0.1$. A possible relation of short-range magnetic order to pseudogap formation was suggested and actively discussed \cite{PseudogapSF,PseudogapSad,PseudogapOnufr,PseudogapToschi,PseudogapMag1,PaseudogapMag2,PseudogapMag3,Sachdev1,Sachdev2,BonettiMetzner}. 

To describe the low-energy spin excitations of magnetic systems  
the nonlinear sigma model (NLSM) can be used, see, e.g., Ref. \onlinecite{Auerbach} for a review. The
corresponding excitations in this model are parameterized by the temporal and spatial spin stiffnesses, which can be determined from the microscopic analysis. For insulating magnets, the classical version of the $O(3)$ nonlinear sigma model was
derived in the continuum limit of the classical Heisenberg model, assuming commensurate antiferromagnetic order
\cite{ZinnJustin,NelsonPelkovits}. Later this derivation was also generalized to
the quantum case \cite{Haldane,CHN,CSY}. The  nonlinear sigma
models in the $O(3)/O(2)$ (Refs. 
 \onlinecite{DombreRead,Mouhanna}) and $CP^{N-1}$ (Refs. \onlinecite{CPN,CSS,Azaria}) manifolds were considered in a continuum limit of frustrated quantum
antiferromagnets with spin spiral ground state.

The situation in itinerant systems is more involved. The derivation of the NLSM, which also allows for obtaining the respective spin stiffnesses, was first 
proposed in Refs. \onlinecite{Weng,Schulz,Dupuis,Dupuis1,Sachdev,Sachdev1} using Hubbard-Stratonovich transformation. This  derivation introduced in the action the effective
fluctuating field, corresponding to the order parameter, which is then fixed at its mean field value. More recently, using the SU(2)-symmetric gauge theory, which does not introduce symmetry
breaking terms or condensation of gauge fields in the action, was proposed for derivation of the nonlinear sigma model and obtaining respective spin stiffnesses
\cite{Bonetti,BonettiThesis,BonettiMetzner,BonettiErratum,OurSecond}. In this approach the symmetry is broken
only at the level of dressed single- and two-particle Green's functions via
spin-asymmetric self-energy and the resulting renormalized interaction
vertices. {The broken spin symmetry in the so-called chargon sector allows one to describe suppression of spectral weight at the points of the Fermi surface close to $(\pi,0)$ and $(0,\pi)$ (see, e.g., Ref. \onlinecite{BonettiMetzner}).}
The respective Ward identities for the SU(2) symmetric approach were derived in Refs. \onlinecite{Bonetti,BonettiThesis,BonettiErratum,OurSecond}. 
It was emphasized in Ref. \onlinecite{OurSecond} that the temporal spin stiffness is in general dynamic, and possesses strong frequency dependence away from half filling. This frequency dependence becomes more pronounced with increase of doping.

In the present paper, we consider the doping- and temperature dependence of spatial and temporal spin stiffnesses of the nonlinear sigma model, obtained within the mean field approach for incommensurate magnetic order of the Hubbard model.
While previous consideration \cite{BonettiMetznerStatic,BonettiNew}  was restricted to the static limit, we uncover the role of the dynamic effects in temporal spin stiffness.
{We emphasize that strong frequency dependence, which occurs already in the low-energy regime, corresponds to consideration of gapped magnetic modes, which are absent in the standard spin-wave approaches, but were discussed recently within the bond-operator technique \cite{Syromyatnikov}}. Using the obtained results, we calculate the temperature- and doping dependence of correlation length and show that {the gapped modes are crucial to obtain} qualitative agreement with the experimental data on high-$T_c$ compounds. 






{\it Magnetic correlations.}
{We consider magnetic correlations in a generic electronic system with
spiral magnetic order in the $x,z$ plane.
For the reasons described below, we consider mainly the out of plane {magnetic} susceptibility in Matsubara formalism $\chi^{yy}_{\mathbf{Q}+\mathbf{q},\omega_n}=\bar{\chi}^{yy}_{\mathbf{Q}+\mathbf{q},\omega_n}=
{\int d\tau e^{i\omega_n \tau} \langle S_{\mathbf q}^y(\tau) S_{\mathbf -q}^y(0)\rangle}$ ({$S^y_{\mathbf q}$ is the Fourier transform of $y$-component of the itinerant spin operator}, the bar denotes the local spin frame, $\tau$ refers to imaginary time). This susceptibility depends on temporal and spatial 
spin stiffnesses $\chi_{\rm{op},\omega_n}$,$\rho_{\rm{op},xx/yy}$ as parameters, which can be determined from microscopic analysis, and characterize the low energy magnetic excitations \cite{Bonetti,BonettiErratum,BonettiThesis,BonettiMetzner,OurSecond}, 
\begin{equation}
    \chi^{yy}_{\mathbf{Q}+\mathbf{q},\omega_n} = \frac{m^2}{\Delta + \chi_{\rm{op},\omega_n} \omega_n^2 + \rho_{\rm{op},x} q_x^2 + \rho_{\rm{op},y} q_y^2},
    \label{out_of_plane_response}
\end{equation} 
where $m$ is the {mean-field} long-range order parameter.
{The actual} short-range magnetic order is characterized by spin gap $\Delta$, determining the correlation length $\xi$.}


To obtain correlation length, we consider the leading order of $1/N$ expansion in the massive CP$^{N-1}$ model 
\cite{CPN,CSS,BonettiMetzner,SM}. 
 In this case, only the out-of-plane mode of magnetic excitations contributes to the correlation length, describing magnetic excitations in the large number of components $N$ limit. 
The 
value of an energy gap $\Delta$
is determined in this approach by the respective sum rule \cite{CSS,SM}
\begin{equation}
    T \sum_{\omega_n} \iint \frac{d^2 q}{(2 \pi)^2} \; \chi^{yy}_{\mathbf{Q}+\mathbf{q},\omega_n} = \frac{m^2}{2}.
    \label{full_gap_equation}
\end{equation}
where the integration is carried over sufficiently small momenta $\mathbf{q}$  
(the chosen integration region is specified in Supplemental Material\cite{SM}),
the factor 1/2 in the right-hand side accounts for the presence of two non-identical magnetic wave vectors $\pm {\bf Q}$ in the Brillouin zone.
{In the polar coordinates {the}} equation for $\Delta$ {reads}
\begin{equation}
    1=T\sum\limits_{\omega_n} \int\limits_{-\pi}^{ \pi} \frac{d \phi }{
    4 \pi^2 \,\rho(\phi)} \ln{ \left( 1 + \frac{q_{ m}^2 (\phi) \rho(\phi)}{\Delta + \chi_{\rm{op},\omega_n} \omega_n^2} \right)},
    \label{gap_equation}
\end{equation}
where $\rho(\phi) = \rho_{\rm{op},x} \cos^2{(\phi)} + \rho_{\rm{op},y} \sin^2{(\phi)}$, {and  $q_{ m} (\phi)$ is the maximum bound of $q$-integration, see Supplemental Material\cite{SM}}. We interpret contributions from finite frequencies $\omega_n$ as an impact of dynamic spin fluctuations on correlation length.

Good quantitative description of an energy gap at low and intermediate temperatures can be obtained 
by separating the frequency sum into the possibly singular $\omega_n = 0$ and finite frequency contributions and approximate the latter contribution, which is finite for 
$\Delta=0$, as an integral over frequencies. In the limit of small spin gap, we find the equation 
\begin{equation}
    \Delta(T) = \bar{q}_m^2 \bar{\rho}{e^{{-
    2\pi\kappa(T) \bar{\rho} (T)}/T} 
    } \label{delta_T}
\end{equation}
with spatial spin stiffness renormalization factor $\kappa(T)$ defined as
\begin{equation}
    \kappa(T) \equiv  1-\frac{1}{
    \pi  \bar{\rho} }\int\limits_{{\pi T}}^{+\infty}\frac{d \omega}{2 \pi} \int{\frac{d\phi}{2\pi}} \ln{ \left( 1 + \frac{q_{ m}^2(\phi) \bar{\rho}}{\Delta(T) +  \chi_{\rm{op},\omega} \omega^2} \right)},
    \label{DeltaT}
\end{equation}
where we characterize the spin gap by the averaged spatial spin stiffness $\bar{\rho} = (\rho_{\rm{op},x}\rho_{\rm{op},y})^{1/2}$, since spin stiffnesses $\rho_{x,y}$ are close to each other,  and perform analytic continuation of the susceptibility $\chi_{{\rm op},\omega}$ to the whole imaginary axis, $\bar{q}_m^2=\exp(\int d\phi \ln q_m(\phi))/\pi$. Note that in general spatial spin stiffness $\bar\rho
$ is temperature-dependent. This fact is of extreme importance at intermediate temperatures due to the proximity to phase transitions where spatial spin stiffness vanishes. Apart from this effect, temperature dependence of $\kappa(T)$ comes from two different sources: lower frequency integration bound $\pi T$ and  temperature dependence of the energy gap $\Delta(T)$. Both significantly reduce the role of quantum fluctuations at finite temperatures. The corresponding correlation length can be estimated as $\xi(T)=\sqrt{\bar{\rho}/\Delta(T)}$.

There are two distinct regimes of
{spin gap temperature dependence}
with lowering temperature. If we assume that $\Delta\to0$ for $T\to0$ than $\kappa(T) \to \kappa(0) \equiv \kappa > 0$ 
{and}
the equation for the spin gap simplifies to
\begin{equation}
    \Delta(T) \approx \bar{q}_m^2 \bar{\rho} e^{-2
    \pi\kappa\bar{\rho}/T}.
    \label{xi_exp}
\end{equation}
where
\begin{equation}
    \kappa = 1-\frac{1}{
    \pi  \bar{\rho} }\int\limits_{0}^{+\infty}\frac{d \omega}{2 \pi}\int{\frac{d\phi}{2\pi}}  \ln{ \left( 1 + \frac{q^2_{ m}(\phi) \bar{\rho}}{\chi_{{\rm op},\omega} \omega^2} \right)}
\end{equation}
In this case, the role of dynamic quantum fluctuations is only in the renormalization of spatial spin stiffness $\bar{\rho} \to \kappa \bar{\rho}$ in the exponent, similarly to the previous analysis of $O(N)/O(N-1)$, Ref. \onlinecite{CSY} and massive $CP^{N-1}$, Ref. \onlinecite{CSS}  nonlinear sigma models. In this regime according to the equation (\ref{xi_exp}) we define the crossover temperature $T^*=2 \pi \kappa \bar{\rho}$ to the quasi-long range order state (denoted also as the renormalized classical regime \cite{CHN}), such that at $T<T^*$ the correlation length gets exponentially large.

Another possibility for $\Delta(T)$ is to approach a finite value with $T\to 0$. In this regime (denoted also as quantum disordered regime \cite{CHN}), the correlation length reaches a finite value with $T\to0$ and thus the system's ground state is disordered. Finiteness $\Delta(T)$ in the denominator of Eq. (\ref{DeltaT}) is important in this regime; 
the equation for the zero-temperature limit of the gap $\Delta(0)$ can be obtained from Eq. (\ref{DeltaT}) by putting $T=0$ and
$\kappa(T=0) = 0$, which implies that the renormalized spatial spin stiffness is zero in this case. 

\begin{figure*}[t]
    \centering    

\includegraphics{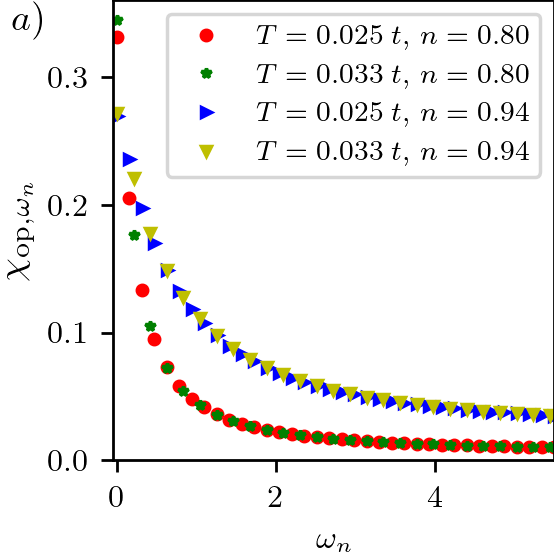}
\hspace{0.1cm}
\includegraphics{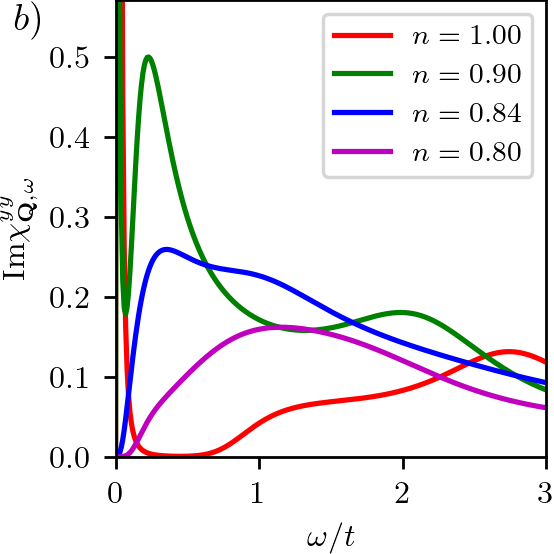}
\hspace{0.1cm}
\includegraphics{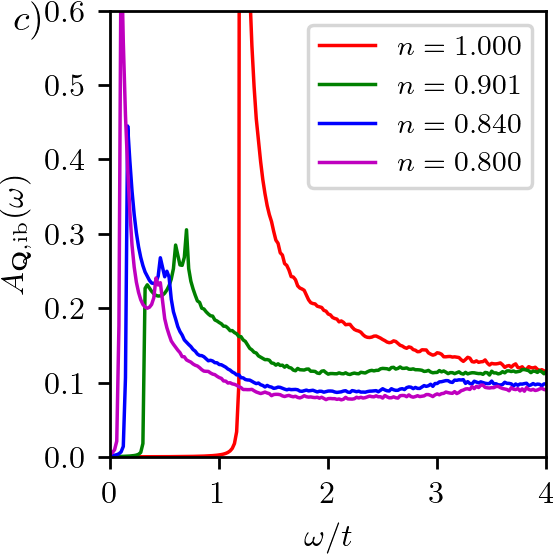}
    \caption{(a) Dynamical temporal spin  stiffness for out-of-plane mode as a function of bosonic Matsubara frequency $\omega_n$ for various temperatures and electron concentrations.
    (b,c) The  real frequency dependence of the imaginary part of the out-of-plane staggered dynamic susceptibility $\text{Im}\chi^{yy}_{\mathbf{Q},\omega}$ (b) and interband spectral weight $A_{{\mathbf{Q}},\rm ib}(\omega)$ (c) for different electron concentrations $n$ at temperature $T=0.025t$.} 
    \label{spin_stiffness_from_n}
\end{figure*}

{Although the above presented analysis parallels that considered previously for the NLSM (see, e.g., Refs. \onlinecite{CHN,CSY}), we emphasize the key difference, which originates from the frequency dependence of the temporal spin stiffness $\chi_{{\rm op},\omega}$. As we show below, it yields strong suppression of magnetic correlations, not allowing}
long range magnetic order to establish at $T=0$ and  leading to the quantum disordered state. {To characterize the respective tendency to short-range magnetic order,} 
we define the crossover temperature of the onset of short-range magnetic order $T_{\rm SRO}$ which fulfills the equation
        $2 \pi \kappa(T_{\rm SRO}) \bar{\rho} (T_{\rm SRO}) = T_{\rm SRO}$,
such that 
{$\xi(T_{\rm SRO})\simeq 1/\bar{q}_m\sim 1$}.
Despite similarity to the definition of the crossover temperature $T^*$, the temperature $T_{\rm SRO}$ has a completely different meaning and denotes the temperature below which (i.e. at $T\ll T_{\rm SRO}$) the gap becomes small, and hence the correlation length is essentially large ($\xi \gg 1$). 


{{\it Spin stiffnesses within the mean-field approach.}
{{For numerical analysis of magnetic excitations} we consider Hubbard model on the square lattice as a prototypical {model describing} strong magnetic correlations},
\begin{equation}
    H = - \sum_{i, j, \sigma } t_{ij}{c_{i\sigma}^\dagger c_{j\sigma}}+U\sum_i{n_{i\uparrow}n_{i\downarrow}}.
    \label{hubbard_model}
\end{equation}
We suppose the hopping $t_{ij}=t$ between nearest neighbors (which is used as a unit of energy) and $t_{ij}=-t'$ for next-nearest neighbors and
{employ a mean‑field approximation, which leads to a gapped electronic spectrum
with dispersion $E^\pm_{\mathbf k}$. The magnetization $m$ is obtained via solution of the respective mean-field equations; minimization of the free energy allows us to
find ${\mathbf Q}$ as a function of filling and temperature.}  {Spin stiffnesses}
are expressed through the spin- and current correlation functions, {subsequently calculated within the random phase approximation approach}, see Refs. \onlinecite{Bonetti,BonettiMetzner,BonettiThesis,BonettiErratum,OurSecond} and Supplemental Material \cite{SM} {for computational details}}. 

{We choose next-nearest neighbor hopping} $t' / t = 0.16$, which is close to that for La$_2$CuO$_4$, and the relatively weak interaction strength $U / t = 2.5$, since for stronger interaction, the incommensurate state in the mean field approach is unstable at low dopings. In view of the used mean field approach, this interaction should be, however, considered as an effective screened interaction, which is typically much smaller than its bare value, cf. Refs. \onlinecite{KataninYamase,IrkhinIgoshev}. The considered interaction also coincides with that used previously in Refs. \onlinecite{BonettiMetznerStatic,BonettiThesis}; it is also close to that 
obtained for not too small doping range in Ref. \onlinecite{BonettiMetzner}.




The obtained frequency dependence of the temporal spin stiffness (see Fig. \ref{spin_stiffness_from_n}a) 
is strong  already for $\omega_n\ll 1$ (in units of $t$), where the NLSM is applicable. 
While the static limit varies only weakly with electron concentration $n$, the decrease of the high-frequency asymptote (which is also almost temperature-independent) corresponds to the strengthening of quantum fluctuations with lowering $n$. These quantum fluctuations are related to the finite spectral weight at finite energies. In Fig. \ref{spin_stiffness_from_n}b we show the resulting frequency dependence of the out-of-plane susceptibility at the real frequency axis, obtained by analytic continuation. At and close to half filling we find the peak at low frequencies corresponding to a vanishing or small spin gap. However, apart from that, we observe generally two other peaks.
One can see that the position of the lower of the two peaks coincides with the low-energy edge of the inter band spectral function     $A_{{\mathbf{Q}},\rm ib}(\omega)=\sum_{\mathbf{k}} \delta( E^+_{\mathbf{k}+\mathbf{Q}} - E^-_{\mathbf{k}} - \omega) ( f(E^-_{\mathbf{k}}) - f(E^+_{\mathbf{k}+\mathbf{Q}}))$, shown in Fig. \ref{spin_stiffness_from_n}. The low-energy edge of $A_{{\mathbf{Q}},\rm ib}$ is determined by an indirect gap in the electronic spectrum.
 This peak can be therefore identified with the Higgs mode \cite{Sachdev1,Sachdev2,Sachdev,SachdevHiggs1}. On approaching the paramagnetic mean-field solution (i.e. at substantial doping levels) this peak shifts to very low energies, and therefore, becomes suppressed by the spin gap. Therefore, at sufficiently large dopings, we observe a single peak in the spin spectral functions. These peaks in the spin spectral function remind us also of the gapped magnetic excitations, discussed recently for the square-lattice Heisenberg model within the bond-operator technique \cite{Syromyatnikov}.

{The spatial spin stiffness vanishes at the electron concentration, corresponding to the transition from the antiferromagnetic to the spiral state \cite{OurSecond,BonettiNew} (see also Supplemental Material \cite{SM}). This concentration gradually shifts toward half-filling as temperature decreases
{yielding a}
rapid increase of spin stiffness near half filling, which, being continuous at a finite temperature, converges to a discontinuous function at $n=1$ in the zero-temperature limit $T\to 0$. In this limit, the spatial spin stiffness vanishes when the electron concentration approaches half-filling from the hole-doped side, while remaining finite when approached from the electron-doped side.}


\begin{figure}[t]
    \centering
    \includegraphics{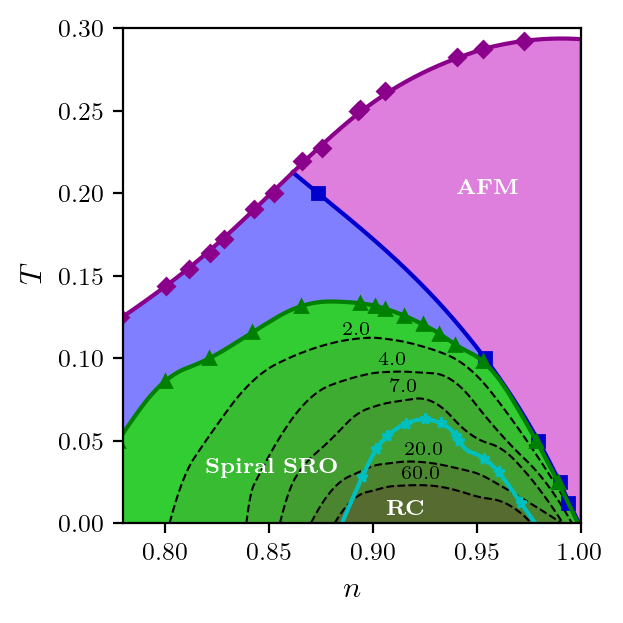}
    \caption{The obtained phase diagram. $T_N^{MF}$ (rhombs)  denote the mean-field transition temperatures to the paramagnetic state, $T_{\text{inc}}$ (squares) mark the transition temperatures between spiral and antiferromagnetic (AFM) states in the mean field approach. $T_{\text{SRO}}$ (triangles) represent the temperatures at which the transition to the paramagnetic state with short range magnetic order, denoted as Spiral SRO and defined by the condition $\xi(T_{\text{SRO}}) > 1$, occurs in the NLSM approach. The dashed lines show contour levels of various values of correlation length (denoted by numbers at the lines in units of the lattice parameter) in the NLSM approach  inside the SRO regime. $T^*$ corresponds to the crossover temperature to the renormalized classical (RC) regime with the exponentially large correlation length in the NLSM approach; the corresponding ground state shows a tendency to long-range order. {The violet region possesses spiral order in the mean field approach, but has small correlation length $\xi\lesssim 1$ in the NLSM approach.}}
    \label{phase_diagram1}
\end{figure}



{\it Phase diagram.} {In Fig. \ref{phase_diagram1}, we present the resulting phase diagram which shows crossover temperatures to the state with short- ($T_{\rm SRO}$) and quasi-long range ($T^*$) magnetic order, as well as regions where magnetic order is entirely destroyed by quantum fluctuations. 
The phase boundaries $T^{MF}_N$ and $T_{\text{inc}}$ represent the transitions between different long-range ordered states, as obtained within the pure mean-field framework. }
{Near the transition between spiral and antiferromagnetic states, the long-range magnetic order is destroyed by purely static magnetic fluctuations, which arise due to the vanishing of the spatial spin stiffness, $\bar{\rho} \to 0$ (see also Ref. \onlinecite{BonettiNew}). 
This effect is absent on the electron-doping side of the phase diagram (not shown here), as the spatial spin stiffness does not vanish in that region.

Away from half-filling, the spatial spin stiffness increases, while the temporal spin stiffness at finite bosonic frequencies becomes suppressed, resulting in enhanced quantum fluctuations. At moderate doping these quantum fluctuations destroy long-range magnetic order at $T=0$, giving rise to states with short-range magnetic order characterized by a correlation length larger than the lattice parameter, $\xi > 1$. These states are located in the green region of the phase diagram in Fig.~\ref{phase_diagram1}.

At the same time, the delicate interplay of the above described effects leads to the interval of electron concentrations where the ground state exhibits long-range magnetic order that is bounded by points of vanishing of the crossover temperature $T^*$. 
The states with finite crossover temperature $T^*$ tend to form an ordered ground state with decreasing temperature. We emphasize in the following that the absence of long-range order at sufficiently large doping is entirely due to the account of the frequency dependence of temporal spin stiffness. 

\begin{figure}[t]
     \centering
     \includegraphics{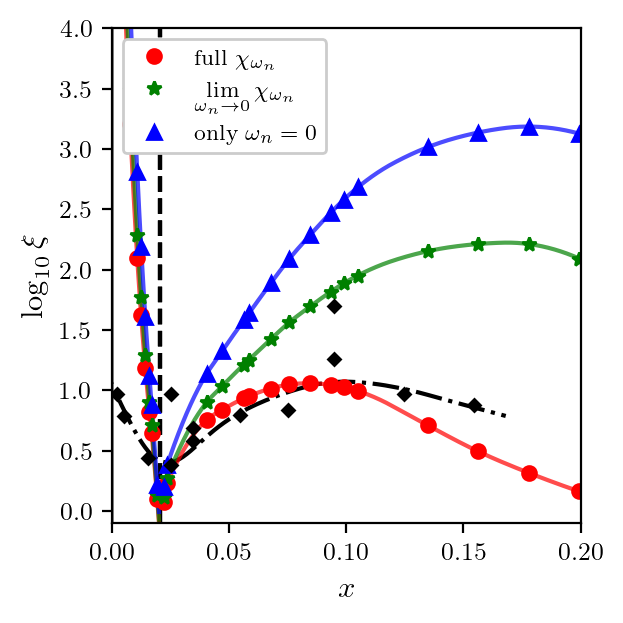}
     \caption{The correlation length $\xi$  (in logarithmic scale) as a function of hole doping $x = 1 - n$ for the temperature ${T} =0.05t$. Red circles represent data obtained using the full dynamical temporal susceptibility $\chi_{\omega_n}$, while green stars correspond to the static approximation applied at all finite bosonic frequencies and blue triangles correspond to the usage of only zero frequency $\omega_n = 0$. The dashed vertical line indicates electron doping at which the phase transition between spiral and antiferromagnetic magnetic orders occurs. Black diamonds and their black dash-dotted approximation represent the experimental data  \cite{ExpInel1,ExpInel2} on La$_{2-x}$Sr$_x$RuO$_4$ compound, expressed in units of the lattice constant $a=\SI{5.36}{\angstrom}$, and shifted horizontally to adjust the point of AFM to incommensurate state transition.}
     \label{xi_from_n}
 \end{figure}

{\it Correlation length.}  In order to examine the role of dynamical quantum fluctuations affecting the system via finite bosonic frequencies and dynamical temporal spin stiffness $\chi_{\rm{op},\omega_n}$, we consider the dependence of the magnetic energy gap, obtained from the equation (\ref{full_gap_equation}), on electron concentration \( n \).  In Fig. \ref{xi_from_n} we compare results of the three different methods for $T = 0.05t$: considering only the zeroth bosonic frequency \(\omega_0\), calculation of the frequency sum with \(\chi_{\omega_n} \approx \lim\limits_{\omega_n \to 0} \chi_{\omega_n}\), and with the dynamic \(\chi_{\omega_n}\) approximated at high frequencies. One can see that the correlation length is decreased by almost an order of magnitude when including finite bosonic frequencies in the sum of Eq. (\ref{full_gap_equation}). Using only the static limit of the temporal spin stiffness in this sum is, however, insufficient; suppression of the dynamic stiffnesses at high frequencies results in a further strong decrease of the correlation length. This effect of quantum fluctuations becomes strengthened with the decrease of temperature (see Supplemental Material \cite{SM}), {yielding an ordered ground state when using the static limit of temporal spin stiffness.} Therefore, we find
%
that the frequency dependence of temporal spin stiffness is of crucial importance for describing spin correlations within the nonlinear sigma model. 

\vspace{-0.05cm}

In Fig. \ref{xi_from_n} we present also the comparison with the experimental data on doping dependence of correlation length in La$_{2-x}$Sr$_x$CuO$_4$. Choosing relatively high temperature $T=0.05t$ in theoretical result is dictated by the necessity of having AFM to incommensurate transition at substancial doping in accordance with the experimental data; is can be justified as modeling the effect of structural disorder and local correlations in the proximity of the metal-insulator transition, which are beyond the mean-field scheme. In particular, our previous study \cite{OurFirst} demonstrated that strong electronic correlations can induce a redistribution of spectral weight between the lower and upper Hubbard bands. This redistribution can prevent the formation of a metallic state, thereby potentially stabilizing the antiferromagnetic ground state even at finite hole doping.}
For comparison, we also shift the experimental data by the doping $\Delta x=-0.025$, to adjust this transition at the same doping, as in the theoretical result. {This difference of dopings may be also an effect of the presence of oxygen states. As noted in Ref. \onlinecite{HoleDoping}, hole carriers may initially occupy nonmagnetic-energy bands rather than directly populating the antiferromagnetic band. 
When considering full frequency dependence of temporal spin stiffness, we find qualitative agreement with the experimental data.

 In summary, we have analyzed the effect of the frequency dependence of temporal spin stiffness on magnetic order in the doped Hubbard model. We find that within the mean-field approach this frequency dependence is crucial to obtain short range magnetic order with a sufficiently small correlation length. In particular, the gapped energy modes, one of which corresponds to the Higgs mode, carry substantial spectral weight, which yields the decrease of correlation length, determined by the sum rule. As a result, we find long-range magnetic order only in the narrow domain of dopings $x\simeq 0.1$, in agreement with the experimental data. We expect that this region of long-range order may be further shrunk by fluctuations, and/or partly unstable with respect to the non-uniform patterns, see, e.g., recent study of Ref. \onlinecite{MetznerFiniteLattice}, which finds stripe instability in approximately the same doping range. 
 

 For future studies, it would be interesting to extend the considered approach to include dynamic local fluctuations within dynamical mean field theory for incommensurate phase \cite{DMFT_Incomm_Licht,DMFT_Incomm,DMFT_Incomm1,SOurFirst,OurSecond}, which may allow considering realistic interaction strengths. Consideration of the effects of Harris disorder within the Yukawa-Sachdev-Ye-Kitaev model  \cite{Harris} is also of certain interest. 

{\it Note added}. Recently, we became aware of Ref. \onlinecite{BonettiNew}, which considered the temperature- and concentration dependencies of spin stiffnesses and correlation length. However, only static temporal spin stiffness was considered in that study.

{\it Acknowledgements}. The calculation of correlation length and phase diagram is supported by the Russian Science Foundation (Grant No. 24-12-00186). {I. A. G. also acknowledges financial support of the derivation of equations for correlation length from BASIS Foundation (Grant No. 24-1-5-152-1).}

\clearpage
\appendix
\widetext

\renewcommand\theequation{S\arabic{equation}}
\renewcommand\thesubsection{\arabic{subsection}}
\renewcommand\thefigure{S\arabic{figure}}
\setcounter{equation}{0}
\setcounter{figure}{0}
\setcounter{page}{1}
\section*{Supplemental Material \\
\lowercase{to the paper }``F\lowercase{requency dependence of temporal spin stiffness and short-range magnetic order\\
in the doped two-dimensional} H\lowercase{ubbard model}"}
\vspace{-0.4cm}
\centerline{I. A. Goremykin and A. A. Katanin}
\vspace{0.5cm} 

\subsection{Connection to the Nonlinear Sigma Model}
\label{nlsm_derivation_section}
In order to describe the long-wavelength magnetic fluctuations on the top of the state with long-ranged magnetic order in chargon sector, one can start from the general action of itinerant electrons on the lattice (described, e.g., by the Hubbard model, Eq. (\ref{hubbard_model}) of the main text)
\begin{equation}
S[c,c^{+}]=\sum_{ij}\int\limits_{0}^{\beta }d\tau \;c_{i}^{+}%
    \left[ \left( \frac{\partial }{\partial \tau }-\mu\right)
    \delta _{ij}-t_{ij}\exp \left( -\vec{r}_{ji}\nabla\right) \right] c_{i}+\int_{0}^{\beta }d\tau H_{\rm int}[c,c^+],
\end{equation}
where $H_{\rm int}[c,c^+]$ is SU(2) symmetric local interaction.
Following Refs. \onlinecite {BonettiMetzner,OurSecond} of the main text, we split the initial fermionic degrees of freedom, described by Grassmann variables $c_i,c^+_i$, into the ``chargon" degrees of freedom $\psi_i,\psi^+_i$, exhibiting long-range magnetic order, and long-wavelength $SU(2)$ rotation field $R$, which introduces spin background distortions at long distances and destroys the long-ranged order, as
\begin{align}
c_i &= {R}_i \psi_i, \, c^+_i = \psi_i^+ {R}_i. \label{Eq2}
\end{align}
Introducing the gauge field $A_{i\mu}$, defined as
\begin{equation}
    A_{i \mu} = \I {R}_i^+ \partial_{\mu} R_i
    \label{A_field_def}
\end{equation}
with $\partial_{\mu} = (\partial_{\tau},\vec{\nabla})$, the action takes the form
\begin{equation}
S[\psi,\psi^{+},A]=\sum_{ij}\int\limits_{0}^{\beta }d\tau \;\psi_{i}^{+}%
\left[ \left( D_{\tau,i} -\mu \right)
\delta _{ij}-t_{ij}\exp ( -\vec{r}_{ji}\vec{D}_{i}) \right] \psi_{i}+\int_{0}^{\beta }d\tau H_{\rm int}[\psi,\psi^+],
\end{equation}
where $D_{\mu,i}=\partial_\mu-{\mathrm i}A_{\mu,i}$. Integrating out fermionic degrees of freedom yields the action of the non-linear sigma model (see, e.g., Ref. \onlinecite{BonettiMetzner} of the main text) 
\begin{equation}
    S^{\text{eff}} [A] = \frac{1}{2}\iint dx\, dx' \,  \mathcal{J}^{ab}_{\mu,x; \nu,x'} A^{a}_{\mu,x} A^{b}_{\nu,x'},
     \label{A_field_action}
\end{equation}
where the components of $A$-field are defined as
\begin{equation}
    A^{a}_{\mu} = \text{Tr} \left[ \sigma^a A_{\mu}\right],
\end{equation}
and $J_{\mu,x;\nu_x}$ is the spin-stiffness tensor, defined by second derivatives of the effective action. To fix the gauge, the infinitesimally small magnetic field should be added (see Refs. \onlinecite{BonettiErratum,OurSecond} of the main text). 

The nonlinearity of the model (\ref{A_field_action}) appears since the gauge fields $A_{\mu,x}$ are subject to the constraint (\ref{A_field_def}) with $R_x^+R_x = 1$.  
The constraint can be automatically satisfied if we express the gauge $A$-field in terms of spinon $z,z^+$ degrees of freedom (see Ref. \onlinecite{BonettiMetzner} of the main text) as
\begin{equation}
    R = 
 \begin{pmatrix}
z_{\uparrow,x} & - z_{\downarrow,x}^* \\
z_{\downarrow,x} & z_{\uparrow,x}^*
\end{pmatrix} R_0,
\label{R_z}
\end{equation}
with the constraint $z^+ z=1$; $R_0$ represents an arbitrary rotation matrix. For the case of spiral magnetic order we choose $
R_0=\frac{1}{\sqrt{2}}
\begin{pmatrix}
e^{\I \pi/4} & e^{\I \pi/4} \\
-e^{-\I \pi/4} & e^{-\I \pi/4}
\end{pmatrix}$. In this representation the gauge field $A$ is expressed in terms of $z$-fields as
\begin{align}
A^x_{\mu} &= 2 \text{Re} (\tilde{z}^+ \partial_{\mu} z), \notag \\  
A^y_{\mu} &= 2 \I z^+ \partial_{\mu} z, \notag \\
    A^z_{\mu} &= - 2 \text{Im} (\tilde{z}^+ \partial_{\mu} z), 
  \label{A_from_z}
\end{align}
(we have also introduced "orthogonal" field $\tilde{z} = \I \sigma^y z^*$).}

\renewcommand\theequation{S\arabic{equation}}
\renewcommand\thesubsection{\arabic{subsection}}

After substitution of these expressions, the action for interacting $z$-field can be written in the form
\begin{equation}
    S[z] = 2\iint dx\, dx' \, \left[ J_{\mu x;\nu x'}^{\text{op}} 
(\partial_{\mu, x} z^+_x) (\partial_{\nu, x'} z_{x'}) 
- J_{\mu x;\nu x'}^{\text{op}} 
(\partial_{\mu, x} z^+_x ) P_{x,x'}  (\partial_{\nu, {x'}} z_{x'}) 
+ J_{\mu x;\nu {x'}}^{\text{in}} \,j_{\mu x} j_{\nu {x'}} \right]
    \label{act}
\end{equation}
where $P_{x,x'}=\hat{1} - \tilde{z}_x \otimes\tilde{z}_{x'}^+$ ($\hat{1}$ is the unity $2\times2$ matrix, $\otimes$ corresponds to the direct product), $j_{\mu,x} = -\frac{\mathbf{i}}{2} \left( \partial_{\mu,x} z^+_x z_x - z^+_x \partial_{\mu,x} z_x\right)$, {$J^{\text{op}}=J^{xx,zz}$ corresponds to the spin stiffness of the out-plane mode, while $J^{\text{ip}}=J^{yy}$  is the in-plane spin stiffness.} The action (\ref{act}) can be extended to the action for $N$-component spinon fields. In the limit $N \to \infty$ due to the completeness relation
\begin{equation}
    z_x \otimes z^+_x + \tilde{z}_x \otimes \tilde{z}^+_x  = \hat{1}
\end{equation}
one can neglect all the terms except for the quadratic part of the action, which results in the standard N-component spinon action. The constraint $z^+ z=1$ yields the equation (\ref{full_gap_equation}) of the main text with the out-of-plane magnetic susceptibility $\chi^{yy}_{\mathbf{q}+\mathbf{Q},\omega_n}$ defined in Eq. (\ref{out_of_plane_response}) of the main text, determined by the saddle-point approximation for the resulting bosonic action. 

\subsection{\lowercase{q}-integration in the sumrule (\ref{full_gap_equation}) and the choice of the cutoff parameter}
\label{q_integration_section}

\begin{figure}[H]
    \centering
    \includegraphics[width=0.6\linewidth]{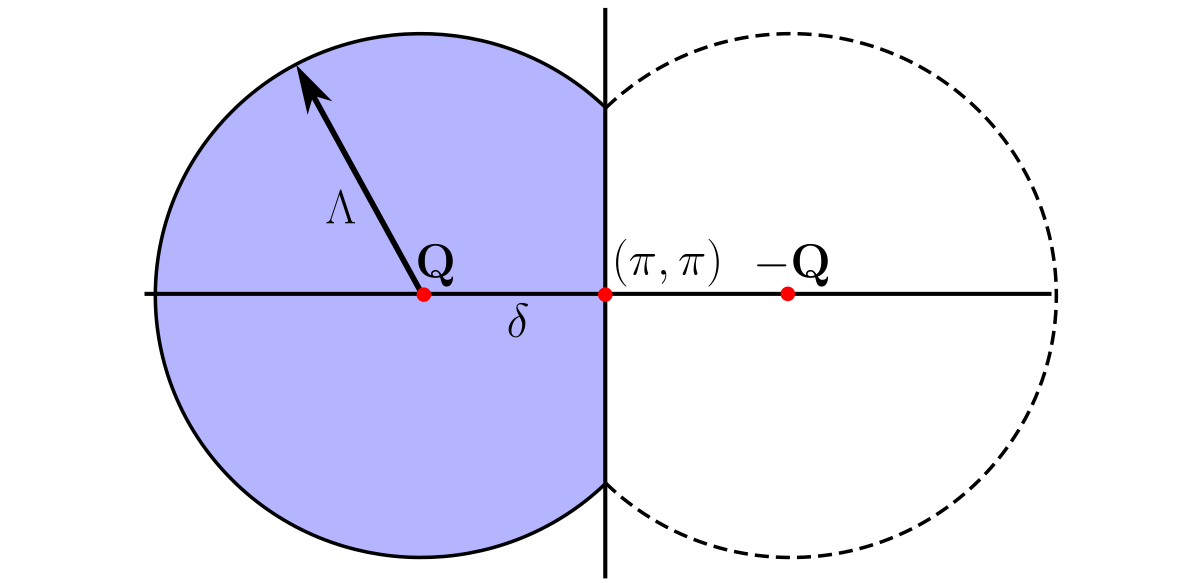}

    \caption{Integration region (shown by colored area) in the sum rule of Eq. (\ref{full_gap_equation}) of the main text}
    \label{figIntegr}
\end{figure}

Since there are generally two points $\mathbf{q}=\pm {\mathbf Q}$ with softening of spin excitations, one should restrict integration in Eq. (\ref{full_gap_equation}) of the main text to the vicinity of one of these points (the presence of two points is accounted for by the factor 1/2 in the right-hand side). 
The respective region of integration is shown in Fig. \ref{figIntegr}. In the radial coordinates, it implies 
\begin{equation}
I=\iint d^2 q=\int\limits_{-\pi}^{\pi} 
d\phi\int\limits_{0}^{q_m(\phi)} q dq,
\end{equation}
where $q_m(\phi)=\delta/\cos\phi$ for $|\phi|<\arccos(\delta/\Lambda)$ and $q_m(\phi)=\Lambda$ otherwise, where $\Lambda$ is a cutoff parameter. In the limit $\delta\rightarrow 0$ (i.e. close to the commensurate phase) we obtain $I=1/2 \iint_{q<\Lambda} d^2 q$, which cancels the factor 1/2 in the right-hand side of Eq. (\ref{full_gap_equation}) of the main text and corresponds to the weight $m^2$ of a {\it single} mode with $q$ close to $(\pi,\pi)$.

The cutoff parameter $\Lambda$ should be chosen in the way that it cuts the short-range degrees of freedom, described by mean-field theory, keeping only the long-range ones. Let us assume for simplicity that the inverse susceptibility in the whole momentum space (bounded as described above) can be approximated as quadratic in momentum, with the respective cutoff $\Lambda_{\rm uv}$. The respective sum rule for the susceptibility reads
\begin{equation}
\frac{\widetilde{S}^{2}}{2}=T\sum_{i\omega _{n}}\iint\limits_{0}^{\;\;\;\;\Lambda _{%
\rm uv}}\frac{d^{2}q}{(2\pi )^{2}}\frac{m^{2}}{\rho q^{2}+\chi \omega
^{2}+\Delta }
\end{equation}
where $\tilde{S}^2$ is the square of magnetic moment. Comparing this to the Eq. (\ref{full_gap_equation}) of the main text, neglecting $\Delta^2$ at $q\sim\Lambda$, and approximating frequency sum by the integral, we find
\begin{equation}
\frac{\widetilde{S}^{2}-m^{2}}{2}=T\sum_{i\omega _{n}}\iint\limits_{\Lambda
}^{\;\;\;\;\Lambda _{\text{uv}}}\frac{d^{2}q}{(2\pi )^{2}}\frac{m^{2}}{\rho
q^{2}+\omega \chi ^{2}}\simeq \frac{m^{2}}{4\pi
\rho }\int\frac{d\phi}{2\pi}\int \frac{d\omega }{2\pi }\ln \frac{q _{\mathrm{uv}}^{2}(\phi)+(\omega /c)^{2}}{q_m
^{2}(\phi)+(\omega /c)^{2}}
\end{equation}%
where $c=(\rho /\chi )^{1/2}$ and $q_{\rm uv}$ is obtained from $q_m$ by replacing $\Lambda\rightarrow \Lambda_{\rm uv}$. For frequency-independent temporal spin stiffness, we obtain%
\begin{equation}
\frac{\widetilde{S}^{2}-m^{2}}{2}\simeq\frac{m^{2}c}{4\pi \rho }{\int\frac{d\phi}{2\pi}}(q_{%
\mathrm{uv}}(\phi)-q_m(\phi) ) \simeq \frac{m^{2}}{4\pi \rho }c(\Lambda _{%
\mathrm{uv}}-\Lambda )
\end{equation}%
From this we find%
\begin{equation}
\Lambda =\Lambda _{\rm uv}-\frac{\widetilde{S}^{2}-m^{2}}{m^{2}}\frac{2\pi \rho 
}{c}
\label{Lambda}
\end{equation}%
Since at $m\rightarrow 0$ we have $\rho\propto \chi\propto m^2$, requiring that $\Lambda$ vanishes at $m\rightarrow
0$ (which implies that all degrees of freedom are described qualitatively
correct by mean-field theory in that case), we find 
$\Lambda _{\rm uv}=%
(\widetilde{S}^{2}/m^{2})\Lambda$ and $\Lambda={2\pi \rho/c }$. Interestingly, the condition $\Lambda \simeq 2\pi \rho /c$ also implies that
the second term in Eq. (\ref{DeltaT}) of the main text is of the order of 
1, i.e. quantum fluctuations become essential for the chosen momentum cutoff. The quantity $2\pi \rho /c$ is in general
temperature- and filling-dependent, but in the considered range $0.8<n<0.95$ it weakly changes with temperature and doping. 
Considering the 
average $\Lambda$, obtained with low- and high-frequency limits of temporal susceptibility, we establish
the  temperature- and filling-independent cutoff  
$\Lambda=0.6$. 

\subsection{Mean field approach}

At the mean field level for {spiral} spin density wave magnetic order with general magnetic wave vector ${\bf Q}$ the Hamiltonian of Eq. (\ref{hubbard_model}) of the main text can be diagonalized and brought to the form \cite{SDW1,SDW2,SDWIc1,SDWIc2,SDWIc3,SDW3,SDW4,SDWOur,OurFirst}
\begin{equation}
    \hat{H} = \sum\limits_{\mathbf{k},\alpha=\pm} E_{\mathbf{k}}^{\alpha} \hat{D}_{\mathbf{k} \alpha}^+ \hat{D}_{\mathbf{k} \alpha} + N U (m^2 - \frac{n^2}{4}),
\end{equation}
where the electronic energy bands are
\begin{equation}
    E^{\pm}_{\mathbf k} = -\tilde{\mu} + \epsilon^{+}_{\mathbf{k}} \pm\sqrt{(\epsilon_{\mathbf{k}}^{-})^2 + \Delta^2_{\text{el}}},\,\,\,
    \epsilon^{\pm}_{\mathbf{k}} = \frac{1}{2}(\epsilon_{\mathbf{k}+\mathbf{Q}/2} \pm \epsilon_{\mathbf{k}-\mathbf{Q}/2}),
    \label{epc_pm}
\end{equation}
$\tilde{\mu} = \mu - {U n}/{2}$,
and the (direct) energy gap in the electronic spectrum is $\Delta_{\text{el}} = U m + {h}/{2}$.

\begin{figure}[t]
    \centering

    \includegraphics{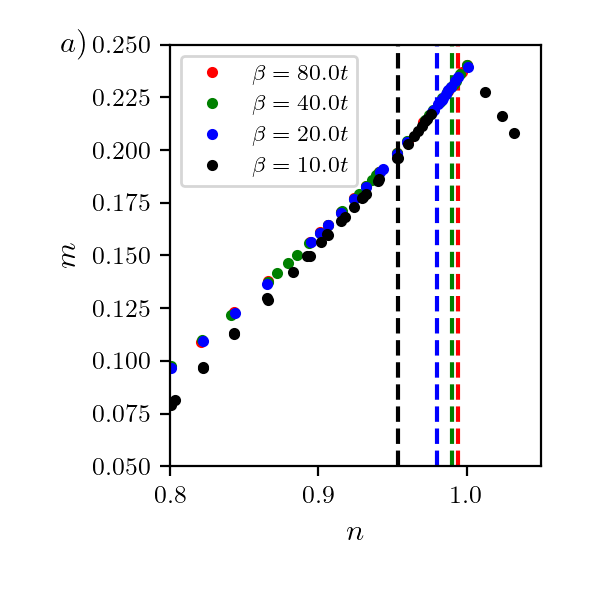}
    \includegraphics{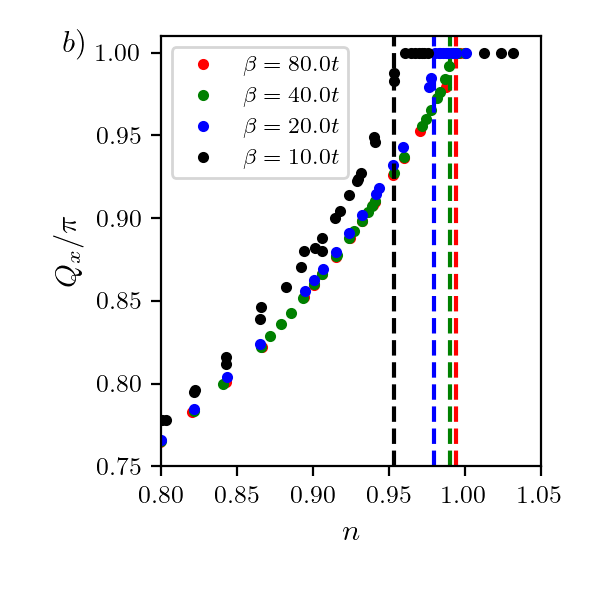}
    \caption{The dependence of a) local magnetization $m$ and b) normalized magnetic wave vector component $Q_x / \pi$ from electron concentration $n$ for different temperatures $T=\frac{1}{\beta}$. Dashed vertical lines correspond to the critical electron concentration $n_c$ of the transition between states with AFM and spiral long-range magnetic order.}
    \label{fig_phase_transitions}
\end{figure}

In order to find thermodynamically stable magnetic phases, we have minimized the free energy potential with respect to the magnetic wave vector $\mathbf{Q}$. The explicit form of the free energy potential $W$ at zero external magnetic field is given by
\begin{equation}
    W = -T \ln Z = -T \sum\limits_{\mathbf{k\alpha}} \ln \left( 1 + e^\frac{-E^{\alpha}_{\mathbf{k}}}{T} \right)  + N U (m^2 - \frac{n^2}{4}).
    \label{free_energy}
\end{equation}
Stable states with long-range magnetic order were found by self-consistent solution of the mean-field equations 
\begin{align}
    & 1 = \frac{U}{2}\sum_{\mathbf{k},\alpha=\pm} \frac{-\alpha f(E^\alpha_{\mathbf k})}{ \sqrt{(\epsilon_{\mathbf{k}}^{-})^2 + \Delta^2_{\text{el}}}}, \,\,\,
    n = \sum_{\mathbf{k},\alpha=\pm} f\left(E_{\mathbf k}^\alpha\right) 
    \label{mean_field_equation}
\end{align}
followed by the minimization of Eq. (\ref{free_energy}) with respect to the magnetic wave vector $\mathbf{Q}$.

The computed dependence of local magnetization $m$ for states with long-range magnetic order and corresponding magnetic wave vector's $\mathbf{Q}$ component $Q_x$ are presented in Fig. \ref{fig_phase_transitions}. Temperature dependence of both quantities is relatively weak, except for the continuous shift of critical doping level $x_c$ of spiral-to-AFM transition to its zero temperature value $x_c = 0$. There is a gradual shift in the critical electron concentration, corresponding to the phase transition between spiral and AFM magnetic orders, towards half-filling as the temperature decreases. 
The obtained order and incommensurability parameters coincide with the results presented in Ref. \onlinecite{BonettiThesis} of the main text.

\subsection{Calculation of spin susceptibility and spin current correlation functions}

From Eq. (\ref{out_of_plane_response}) of the main text, it follows that the spatial spin stiffness can be directly obtained as $\chi_{\rm{op},\omega_n}=  {m^2}/({\bar{\chi}^{yy}_{\mathbf{Q},\omega_n} \omega_n^2})$. 
For the computation of the nonuniform dynamical spin susceptibility in the local coordinate frame, $\bar{\chi}^{\alpha,\beta}_{\mathbf{q}}$, we employ the random phase approximation (RPA), following the expression
\begin{equation}
\bar{\chi}_{\mathbf{q}} = (\hat{I} - \bar{\chi}^{(0)}_{\mathbf{q}} \hat{U})^{-1} \bar{\chi}^{(0)}_{\mathbf{q}},
\label{local_susceptibility}
\end{equation}
where all quantities are treated as matrices in the spin space, indexed by $\alpha, \beta = x, y, z, 0$. Here, $\hat{U}$ represents the inter-electron interaction matrix
\begin{equation}
\hat{U} =
    \left(
\begin{matrix}
  0 & 0 & 0 & -U \\
  0 & U & 0 & 0 \\
  0 & 0 & U & 0 \\
  -U & 0 & 0 & 0
\end{matrix} \right),
\end{equation}
while $\bar{\chi}^{(0)}_{\mathbf{q}}$ denotes the bare susceptibility bubble, defined as
\begin{equation}
    \chi^{(0) \; \alpha\beta}_{ q,\omega_n}=-\frac{T}{4}\sum\limits_{k} \mathrm{Tr}[\sigma^\alpha G_{k} \sigma^\beta G_{k+q} ]
    \label{chi_bubble_expr}
\end{equation}
with $\sigma^\alpha_{\sigma,\sigma'}$ being Pauli matrices, with spin-projection indices $\sigma,\sigma'=\uparrow,\downarrow$ and 
\begin{equation}
G_{\mathbf{k},\nu}^{\sigma \sigma'} =
    \begin{pmatrix}
    i\nu + \mu - U n_{\downarrow} - \epsilon^{+}_{\mathbf{k}} &
    \I \epsilon^{-}_{\mathbf{k}} \\
    -\I \epsilon^{-}_{\mathbf{k}} & i\nu + \mu - U n_{\uparrow} - \epsilon^{+}_{\mathbf{k}}
    \end{pmatrix}^{-1}
\end{equation}
being the one-particle Green functions in local coordinate frame; $\epsilon_{\mathbf k}^\pm$ is defined according to (\ref{epc_pm}).

The susceptibility $\bar{\chi}_{\mathbf{q}}$ obtained from Eq.~(\ref{local_susceptibility}) serves as a key quantity for evaluating the dynamical spin stiffness $\chi_{\text{op},\omega_n}$ and analyzing the stability of mean-field solutions against the emergence of inhomogeneous phases by examining the spectrum of $\bar{\chi}_{\mathbf{q}}$ throughout the entire Brillouin zone.

The spatial spin stiffness of the out-of-plane $\rho_{\text{op},\mu}$ and in-plane $\rho_{\text{ip},\mu}$ modes with $\mu = x, y$ are related to the full gauge kernel function $M^{xx}_{\mathbf{q},\omega_n}$ in the mean-field approximation can be expressed as (Refs. \cite{OurSecond,BonettiThesis,Bonetti,BonettiMetzner} of the main text)
\begin{align}
\rho_{\text{op},n} &= 2 \left({K}_{0,nn}^{(0),xx}+{K}_{0,nn}^{d}\right),\label{rhoop}\\
\rho_{\text{ip},n} &= {K}_{0,nn}^{(0),yy}+{K}_{0,nn}^{d}
,\label{rhoip}
\end{align}
where $K^{d}_{\mu\nu}$ accounts for the diamagnetic contribution, and ${K}_{0,nn}^{(0),xx}$ is a bubble contribution to the correlation function $\langle \langle \hat{J}_{\mathbf{q},{\mu }}^{\alpha }|\hat{J}_{-\mathbf{q},{\nu }}^{\beta }\rangle \rangle$ of spin currents $\hat{J}^{\alpha}_{\mathbf{q},\mu}$, $\alpha = x, y, z, 0$ ($\alpha = 0$ corresponds to the usual charge current operator), $\mu = x, y$. {In $\rho_{in,n} $ we have also neglected the contribution $\Delta \rho_n$ discussed previously in Ref. \onlinecite{OurSecond} of the main text, originating from mixing of different susceptibility components, as it was found to be numerically small.} Similarly to the equation (\ref{chi_bubble_expr}) one can express ${K}_{0,nn}^{(0),xx}$  and  $K^d_{0,nn}$ in terms of single-particle Green functions
\begin{align}
    {K}_{0,nn}^{(0),xx} &= \sum\limits_{s=\pm} \left[ \bar{K}^{(0),xx} -i s (\bar{K}^{(0),xz} -\bar{K}^{(0),zx}) +\bar{K}^{(0),zz}\right]_{q,nn,ss},\\
    K_{q,\mu\nu}^{(0),yy} &= \bar{K}_{q,\mu \nu,++}^{(0),yy} +\bar{K}_{q,\mu \nu,--}^{(0),00} +%
\bar{K}_{q,\mu \nu,+-}^{(0),y0} +\bar{K}_{q,\mu \nu,-+}^{(0),0y},\\
    \bar{K}^{0;\alpha \beta}_{q;\mu\rho,ss'} &=  T \sum_{k} T^{\mu,s}_{{\mathbf{k},\nu},{\mathbf{q}}} \mathrm{Tr} \left[\sigma^{\alpha} G_{k}\sigma^{\beta} G_{k+q} \right] T^{\rho,s'}_{{\mathbf{k}},{\mathbf{q}}},
\end{align}
where lower indices of all terms in the local coordinate frame are the same for all contributions and was taken out of the square brackets. Current vertices are defined as
\begin{align}
T^{\mu \pm}_{{\mathbf k},{\mathbf q}} &= (T^\mu_{{\mathbf k}-{\mathbf Q}/2,{\mathbf q}} \pm T^\mu_{{\mathbf k}+{\mathbf Q}/2,{\mathbf q}})/{2}\,\,\,&(\alpha,\beta=0,y),\notag\\
T^{\mu \pm}_{{\mathbf k},{\mathbf q}} &= T^\mu_{{\mathbf k}+s{\mathbf Q}/2,{\mathbf q}}\,\,\,&(\alpha,\beta=x,z),
\end{align}
$T^\mu_{{\mathbf{k}},{%
\mathbf{q}}}=({t_{\mathbf{k}}^{\mu} + t^{\mu}_{\mathbf{k} + \mathbf{q}}})/2$
 with velocity components $t^{m}_{\mathbf{k}} = {\partial}\epsilon_{%
\mathbf{k}}/{\partial k^{m}}$. 

Diamagnetic contribution can also be expressed as $K_{\mu\nu} ^{d} =\bar{K}_{ \mu\nu,+}^{d, 00} +\bar{K}_{ \mu\nu,-}^{d, yy}$, where
\begin{equation}
    \bar{K}_{ \mu\nu,s}^{d,ab} = -\delta_{ab}(1-\delta_{\mu 0})(1-\delta_{\nu 0})
\sum_k T^{\mu\nu,s}_{\mathbf{k}} \mathrm{Tr}\left[ \sigma^a G_{k} \right].
\end{equation}


\subsection{Spatial spin stiffnesses}

\begin{figure}[H]

\includegraphics{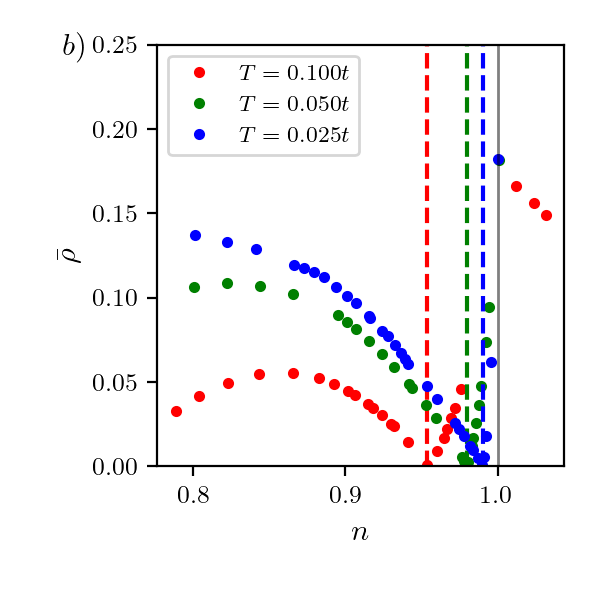}
\includegraphics{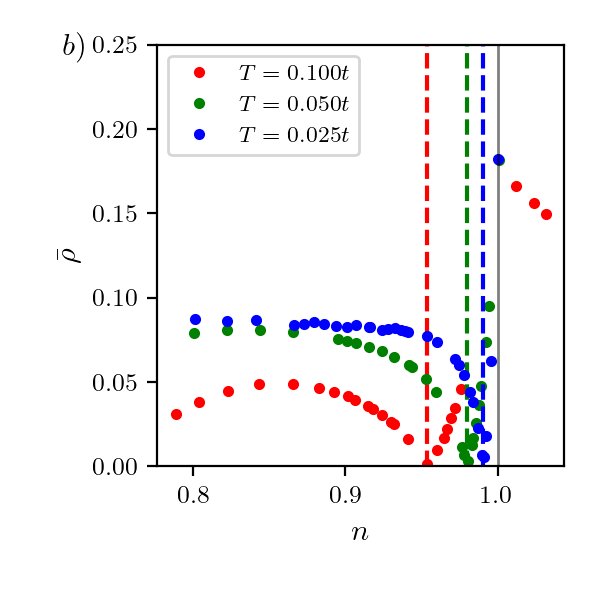}
\caption{The dependence of spatial spin stiffness on electron concentration $n$ for various temperatures $T=\frac{1}{\beta}$ for the out-of-plane (left) and in-plane (right) excitations.}
\label{Figrho}
\end{figure}

For completeness, we show the dependence of spatial spin stiffnesses calculated via equations (\ref{rhoop}) and (\ref{rhoip}) on electron concentration, see  Fig. \ref{Figrho}. The temperature dependence of spatial spin stiffness is different for the cases of out-of-plane and in-plane responses. As temperature decreases the dependence of the in-plane spatial stiffness $\rho_{2,xx/yy}$ becomes sharper in the vicinity of half-filling. With $T\to0$ it converges at all finite doping levels to the value nearly independent of electron concentration and develops a jump at the half-filling. 

\subsection{Frequency dependence of the in-plane temporal spin stiffness}

The frequency dependencies of the in-plane temporal spin stiffness (\ref{local_susceptibility}) for two electron concentrations are presented in Fig. \ref{Figchirhoip}.  Similarly to the out-of-plane susceptibility, discussed in the main text, it decays with the increase of frequency. At low temperatures the frequency spacing becomes finer, enabling accurate extrapolation to obtain the static limit. {We note that while despite strong frequency dependence, $\chi_{\rm{op},\omega_n}$ is continuous at zero frequency for states with spiral magnetic order, i.e. its zero-frequency value coincides with the considered static limit.
{However, the in-plane mode, corresponding to zero momentum transfer in the local coordinate frame, exhibits a discontinuity at zero Matsubara frequency. This discontinuity reflects the difference between the dynamical limit of the retarded susceptibility, $\chi(\omega)$, and the static thermal spin susceptibility, $\chi^{\text{th}} = \chi(\I \omega_n = 0)$. Physically, this difference appears due to conservation of total magnetic moment and the ergodicity of the system's state, {similarly to previous study of the paramagnetic state near Mott transition}  \cite{FrequencyAndJumpIssues}.} The magnitude of the jump is higher at higher doping levels. 
Paradoxically, {\it absence} of the jump in the out-of-plane component yields its faster decay with frequency, since in that case the jump is replaced by smooth, yet rapidly decreased, function. {From the ergodicity point of view, the infinite-time correlations related to the jump {of the in-plane temporal spin stiffness} are replaced by a large but finite correlation time {of the out-of-plane correlations}, manifested as a sharp peak at $\omega=0$ with finite width.} 

\begin{figure}[h!]
    \centering
\includegraphics{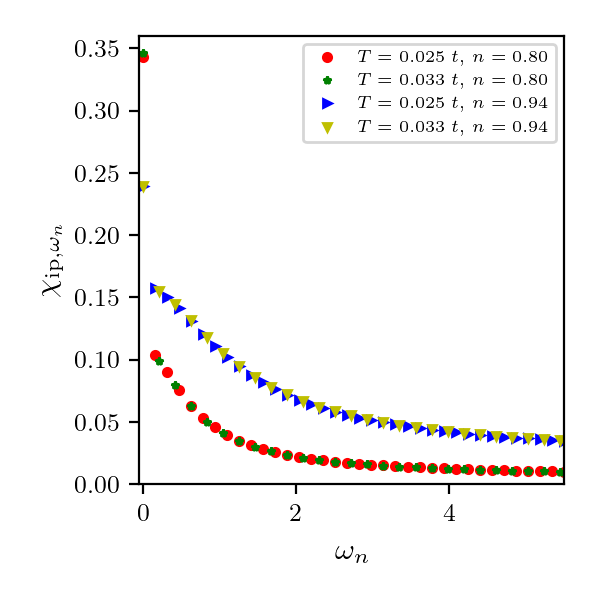}
    
    \caption{Dynamical temporal spin  stiffness for in-plane mode as a function of bosonic frequency $\omega_n$ for various temperatures and electron concentrations.
}
    \label{Figchirhoip}
\end{figure}

\begin{figure}[t]
    \centering
    \includegraphics{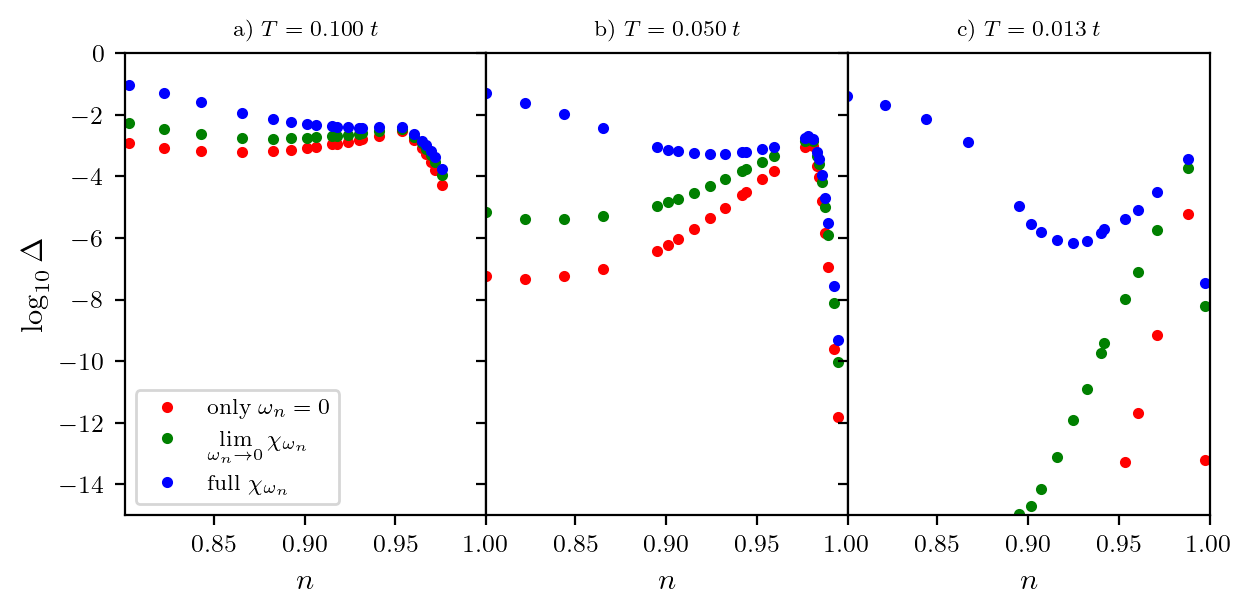}
   
    \caption{The comparison of different approximations for spin gap determined from the Eq (\ref{gap_equation}) of the main text at various temperatures. The dependence of energy gap logarithm $\log_{10} \Delta$ on electron concentration $n$ is computed with inclusion of only zeroth bosonic frequency $\omega_n=0$, all frequencies but with temporal spin stiffness approximated by its static limit $\lim_{\omega_n \to 0} \chi_{\rm{op},\omega_n}$ and the calculation with dynamic temporal spin stiffness $\chi_{\rm{op},\omega_n}$ obtained as a function of bosonic frequency $\omega_n$. 
}
    \label{log_delta_from_n1}
\end{figure}

\subsection{The details of the comparison of the spin gap in different approaches at various temperatures}
\label{appendix_approaches_comparison}

In Fig. \ref{log_delta_from_n1} we present the dependence of the magnetic energy gap $\Delta$ (in logarithmic scale) on electron concentration $n$ for various temperatures. The results are obtained from equation (\ref{full_gap_equation}) of the main text using three different approaches: (i) including only the zero-frequency contribution $\omega_n = 0$, (ii) approximating the dynamic temporal spin stiffness $\chi_{\text{op},\omega_n}$ by its static limit $\lim\limits_{\omega_n\to 0} \chi_{\text{op},\omega_n}$, and (iii) using the exact values of $\chi_{\text{op},\omega_n}$ at all bosonic Matsubara frequencies $\omega_n$. It can be seen that both approaches (i) and (ii) results in exponential decrease of the magnetic energy gap with lowering temperature, although inclusion of finite bosonic frequencies increases $\Delta$ by several orders of magnitude. Full inclusion of temporal spin stiffness frequency dependence results in a completely different trend, that is described in the main text: only for states in a certain region of electron concentrations $n$ around $n \approx 0.9$ does the magnetic energy gap significantly decrease with lowering temperature while approaching a certain limit outside of this stability region.


\bibliographystyleApp{unsrt}





\end{document}